%% file: main.tex
\documentclass[12pt,a4paper,twoside]{article}

\usepackage[english]{babel}
\usepackage[utf8]{inputenc}
\usepackage{csquotes}

\usepackage{microtype}                   
\usepackage[onehalfspacing]{setspace}   
\usepackage{parskip}                    
\usepackage[hang,flushmargin]{footmisc} 

\usepackage[
  top=2cm,
  bottom=2cm,
  left=2cm,
  right=2cm,
  headheight=17pt,
  includehead,includefoot,
  heightrounded
]{geometry}

\usepackage{fancyhdr}
\usepackage[hidelinks]{hyperref}

\usepackage{graphicx}
\usepackage{placeins}     
\usepackage{incgraph}     
\usepackage{rotating}     
\usepackage{float}        
\usepackage{subfiles}     
\usepackage{import}       

\usepackage{booktabs}
\usepackage{makecell}
\usepackage{multirow}
\usepackage{longtable}
\usepackage{varwidth}
\usepackage{array}
\usepackage{xcolor}
\newcolumntype{P}[1]{>{\centering\arraybackslash}p{#1}}

\usepackage{amsmath,amssymb,amsfonts}

\usepackage{natbib}

\usepackage[colorinlistoftodos]{todonotes}

\usepackage{titling}
\pretitle{\vspace{2em}\centering\Huge\bfseries}
\posttitle{\par\vspace{1em}}
\preauthor{\centering\large}
\postauthor{\par\vspace{0.5em}}
\predate{\centering\small}
\postdate{\par\vspace{2em}}

\title{Next-Generation Conflict Forecasting\\Unleashing Predictive Patterns through Spatiotemporal Learning\\\textsc{DRAFT}}
\author{Simon Polichinel von der Maase}
\date{\today}

\begin{document}

\begin{titlepage}
    \centering
    \vspace*{3cm}

    {\Huge \textbf{Next-Generation Conflict Forecasting:\\} Unleashing Predictive Patterns through Spatiotemporal Learning\par}
    \vspace{1.5cm}
    {\Huge Working Paper}
    \vspace{1cm}

    \noindent\makebox[\textwidth]{\Large\textemdash\quad$\diamond$\quad\textemdash}
    \vspace{1cm}

    {\Large Simon Polichinel von der Maase \par}
    \vspace{0.5cm}
    \textit{PRIO -- Oslo} \\
    \textit{\today}
    \vspace{1.5cm}
    
\textbf{Author’s note}: This paper is based on material originally included in the author’s doctoral dissertation \citep{vondermaase2023}. It is currently being revised and extended as a standalone article, drawing on feedback and discussions with colleagues and reviewers.

    \vfill
    \hspace{1cm}
    \includegraphics[width=0.25\textwidth]{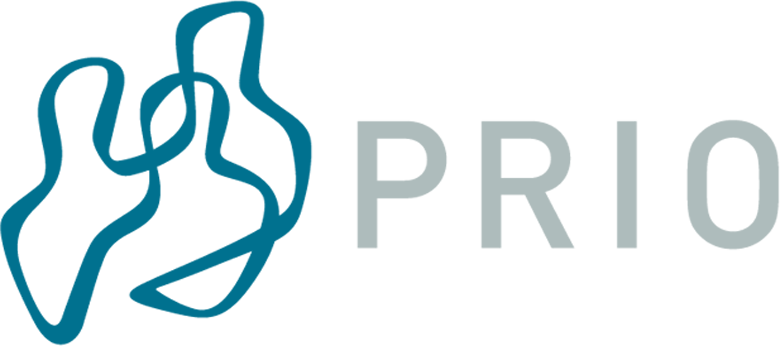}
    \hspace{1cm}

    \thispagestyle{empty}
\end{titlepage}


\input{sections/abstract}


\section{Introduction}
\input{sections/introduction}

\section{Current approaches, challenges, and solutions} 
\input{sections/current_approaches}

\section{Design and development}
\input{sections/design}

\section{Evaluation}
\input{sections/evaluation}

\section{Conclusion}
\input{sections/conclusion}

\section*{Replication data and source code}
\input{sections/replication}

\section*{Acknowledgements}
\input{sections/acknowledgements}

\section*{Bibliography}
\bibliographystyle{apalike}
\bibliography{main.bib}

\section{Appedix}
\input{sections/appendix}

\end{document}

%% file: sections/abstract.tex
\begin{abstract}
Forecasting violent conflict at high spatial and temporal resolution remains a central challenge for both researchers and policymakers. This study presents a novel neural network architecture for forecasting three distinct types of violence -- state-based, non-state, and one-sided -- at the subnational (priogrid-month) level, up to 36 months in advance. The model jointly performs classification and regression tasks, producing both probabilistic estimates and expected magnitudes of future events. It achieves state-of-the-art performance across all tasks and generates approximate predictive posterior distributions to quantify forecast uncertainty.

The architecture is built on a Monte Carlo Dropout Long Short-Term Memory (LSTM) U-Net, integrating convolutional layers to capture spatial dependencies with recurrent structures to model temporal dynamics. Unlike many existing approaches, it requires no manual feature engineering and relies solely on historical conflict data. This design enables the model to autonomously learn complex spatiotemporal patterns underlying violent conflict.

Beyond achieving state-of-the-art predictive performance, the model is also highly extensible: it can readily integrate additional data sources and jointly forecast auxiliary variables. These capabilities make it a promising tool for early warning systems, humanitarian response planning, and evidence-based peacebuilding initiatives.
\end{abstract}
\newpage

%% file: sections/introduction.tex
At the heart of peace research lies the task of forecasting violent conflict \citep{singer1973peace}. Reliable forecasting frameworks, capable of estimating the probability, magnitude, and locations of future conflicts, can be used by policymakers and practitioners to reduce the impact and fallout of conflicts \citep{Ward_Greenhill_Bakke_2010, perry_2013, hegre2017introduction, hegre2017evaluating, hegre2019views, vesco2022united}. In addition, such frameworks can provide researchers with valuable insights into the underlying causes and dynamics of violent conflicts, leading to innovative theories and approaches to prevention, mitigation, and resolution \citep{Schrodt_2014, chadefaux2017conflict, hegre2017introduction, hegre2017evaluating, cederman2017Gurr}.\par

To generate such forecasts, contemporary approaches rely on a diverse array of data sources, including conflict history, protest records, geographic data, socioeconomic indicators, and institutional factors \citep{ward2017lessons, hegre2017introduction, hegre2019views}. To generate predictions, such data are transformed into theoretically motivated representations, which then serve as input features for one or more machine learning models \citep{perry_2013, chadefaux_2014, mueller_2016, hegre2019views, hegre2021views2020}. Over the past decade, this approach has shown considerable promise, leading to increasingly reliable and robust conflict forecasts \citep{hegre2017introduction, vesco2022united}.\par

However, many of these data sources suffer from low-variance signals due to inherent inertia, coarse measurements, or both, making them unsuitable for predicting dynamic developments. As a consequence, data on past conflict patterns, being relatively well-maintained and highly disaggregated, serve as the foundation for today's most powerful predictors \citep{hegre2019views, hegre2021views2020, hegre2022lessons, radford2022high, malone2022recurrent, vesco2022united}.\par

Yet, in this article, I argue that the full predictive potential of past conflict patterns remains largely untapped. First, data on past conflict patterns are usually transformed into theoretically motivated features designed to mirror phenomena such as conflict traps and diffusion, but there is no guarantee that these manual transformations capture the predictive potential of such phenomena effectively. Second, the high predictive power observed in past patterns may not solely arise from such theoretically grounded phenomena. Past conflict patterns might also act as \emph{high-variance predictive proxies} for various relevant but unobserved or poorly measured features. Consequently, manually created features might only capture a fraction of the predictive signal contained in past conflict patterns, thus leading to suboptimal forecasting results.\par


To address this challenge, I present a novel Neural Network architecture, a \emph{Monte Carlo Dropout LSTM U-net} denoted \emph{HydraNet}, which is meticulously designed to automatically learn and extrapolate salient spatiotemporal patterns at a highly disaggregated level of analysis (LoA). By doing so, it greatly alleviates the burden on researchers who would otherwise manually construct hosts of spatiotemporal features. Even more crucially, it guarantees that the learned spatiotemporal patterns are optimized for the specific forecasting task.\par

Notably, the architecture is designed to generate forecasts of multiple targets simultaneously. In the specific implementation presented here, I exemplify this by forecasting three distinct types of violence: state-based, non-state-based, and one-sided violence. This capability not only offers the convenience of employing a single model for addressing a set of interrelated tasks concurrently but also ensures that interrelated patterns between the various targets are learned and leveraged to optimize prediction power.\par

To further enhance the architecture for the specific tasks of conflict forecasting and to showcase its flexibility, the implementation presented here also estimates both the probability and magnitude of future violence -- effectively increasing the number of targets to six. This ensures a comprehensive assessment of potential future conflict scenarios.\par

Lastly, to quantify (some of) the uncertainty inherent to conflict forecasting, the architecture is designed to generate a predictive posterior distribution for each forecasted target. This greatly increases the insights that can be derived from the generated forecast and enhances the capacity to anticipate unlikely but potentially devastating events -- thus providing valuable insights to scholars, practitioners, and policy-makers alike.\par

By combining these advanced features, the presented architecture tackles several central tasks crucial to our endeavor of forecasting future large-scale violence at a highly detailed LoA -- thus reflecting a comprehensive and holistic approach to conflict forecasting.\par

To ensure comparability with the current state-of-the-art, I evaluate the architecture's performance according to the same scheme as \cite{hegre2021views2020}: I use data covering Africa and the Middle East\footnote{Results excluding the Middle East are also provided to align with \cite{hegre2021views2020}.} from January 1990 to December 2019, with the last 36 months used for out-of-sample evaluation. The unit of analysis is monthly, subnational grid cells. The proposed architecture achieves new state-of-the-art performance across all tasks. Importantly, this is accomplished while relying exclusively on past conflict patterns as predictors.\par 

Thus, this paper represents a substantial advancement in the field of conflict forecasting, offering a powerful tool for generating forecasts at highly disaggregated granularity. Its contributions carry practical implications for conflict prevention, mitigation, and resolution while also providing novel insights and methodologies for the study of violent conflicts. Furthermore, the architecture's flexible design holds promise for addressing other tasks exhibiting similar spatiotemporal patterns, extending its applicability well beyond conflict forecasting.\par

%% file: sections/current_approaches.tex
The goal of conflict forecasting is to generate reliable predictions concerning the future developments of violent conflicts, preferably with a highly disaggregated spatiotemporal LoA \citep{hegre2017introduction, hegre2019views, hegre2021views2020, radford2022high}. These forecasts typically include predictions regarding the timing, location, probability, and potential magnitude of future conflict events, with battle-related fatalities conventionally used as an indicator of the presence and magnitude of violent conflict \citep{perry_2013, chadefaux_2014, mueller_2016, hegre2017introduction, hegre2019views, hegre2021views2020}. 

To generate such forecasts, two key components are required: suitable models and relevant features derived from realized data \citep{hegre2017introduction}. The models utilized are often borrowed from statistics and machine learning, with Random Forest and XGBoost being common choices \citep{perry_2013, hegre2019views, hegre2021views2020}. These models excel at capturing nonlinear relationships and interactions in structured datasets but cannot learn complex spatiotemporal dependencies without extensive manual feature engineering.\par 

The features used to train these models are often categorized into various \emph{themes} such as conflict history, protest, geography, and country-level factors. Conflict history features are created to encompass aspects like lagged conflict fatalities, the proportion of months with conflict, and the time since the last conflict. Similarly, protest features involve lagged protest, time since the last protest, and spatial lag of protest. Geography features may include distance to resources, terrain characteristics, land-use types, and distances to capitals and borders. Country-level features can consist of political regime indicators, time since independence or regime change, population demographics, and economic factors \citep{ward2017lessons, hegre2019views, hegre2021views2020}.\par 

Yet, while theoretical and statistical associations between such features and conflict exist, this does not guarantee predictive performance \citep{Goldstone_2010, Ward_Greenhill_Bakke_2010}. Indeed, these features often have two notable shortcomings. First, structural features tend to exhibit inertia and relatively slow development compared to conflict levels, which often manifest as a mixture of persistent long-term trends, rapidly oscillating short-term trends, and contextual idiosyncrasies \citep{chadefaux2017conflict}. Second, even features expected to exhibit adequate variance are often measured at highly aggregated temporal levels spanning years to decades. As such, predicting short-term developments using such features is infeasible \citep{hegre2021can, vesco2022united, hegre2022lessons}.\par

To address this, some researchers have turned to public sentiment. For instance, text data from news outlets has been used to detect shifts in sentiment among elites, the public, and the media, aiming to predict upcoming turmoil \citep{chadefaux_2014, mueller_2016}. However, this approach faces difficulties in achieving spatially disaggregated forecasts due to the focus of news articles on national events, resulting in limited coverage of local developments. Consequently, capturing variations in conflict dynamics across sub-national units becomes challenging.\par

Notably, the most powerful predictors currently at our disposal pertain to a simple theme: conflict history. In other words, the features with the highest prediction power in existing work are simply spatiotemporal transformations of our target, observed conflict fatalities \citep{hegre2019views, hegre2021views2020, hegre2022lessons, radford2022high, malone2022recurrent, vesco2022united}. 

These \enquote{conflict history} features are usually designed to emulate one of two closely connected phenomena: \enquote{Conflict traps} and \enquote{conflict diffusion}. Conflict traps refer to temporal feedback loops where a region becomes trapped in a self-reinforcing cycle of violence that hinders conflict resolution and increases the likelihood of future conflicts \citep{walter2004does, collier2002understanding, raknerud1997hazard, beck1998taking, collier2003breaking, hegre2017evaluating, hegre2021can}. Similarly, self-reinforcing feedback loops of violence can extend through space as well as time, leading to conflict diffusion \citep{buhaug2008contagion, ol2010afghanistan, schutte2011diffusion, crost2015conflict, bara_2017}. Given the predictive power of past conflict patterns, creating features that capture such phenomena effectively is naturally considered a crucial task.\par

The challenge, however, lies in our limited understanding of the underlying data-generating process and the appropriate functional forms for these phenomena. Conflict traps are typically modeled using various decay functions \citep{perry_2013, hegre2017evaluating, hegre2021views2020}, while conflict diffusion is modeled using binary or count indicators of neighboring conflict-affected regions \citep{ol2010afghanistan, weidmann_ward_2010predicting, hegre2019views, hegre2021views2020}. Yet, these phenomena result from a wide range of interconnected factors such as political mobilization, military socialization, and the militarization of local authorities \citep{wood2008social}. Increasingly influential militaries, profiteers of violence, and support from trans-border ethnic kin \citep{hegre2017evaluating, stedman1997spoiler, Cederman_Gleditsch_Buhaug_2013}. Fragmented political economies, social network disintegration, and polarization of social identities \citep{wood2008social}. Challenges related to reintegrating veterans, firearm circulation, and inter-group grievances \citep{humphreys2007demobilization, lock1997armed, hegre2017evaluating}. Destruction of infrastructure, incurred debt, disrupted trade, impeded growth, and reduced state capacity \citep{slantchev2012borrowed, bayer2004effects, Fearon_Laitin_2003, Collier_Hoeffler_2004}. All factors that can also serve to diffuse conflict through space \citep{buhaug2008contagion, schutte2011diffusion, crost2015conflict, hegre2017evaluating, bara_2017}. Thus, phenomena such as conflict traps and diffusion might well be too complex to capture with relatively simple transformations such as decay functions and lags.\par

To compound the issue, there is no guarantee that the predictive power of past conflict patterns stems solely from theoretically grounded phenomena such as conflict traps and diffusion. Indeed, the nature of the data itself potentially plays a substantial role: Since many conflict-related factors are challenging or infeasible to measure at a sufficiently disaggregated granularity, past conflict fatalities end up serving as fine-grained, \emph{high-variance predictive proxies} for a host of unobserved or inadequately measured factors. E.g., changing strategic incentives, temporary breakdown of local authorities, sudden increased ethnic or religious tensions, abruptly decreased access to resources or services, ceasefires and breakdowns of such, etc.\footnote {I thank Nils B. Weidmann for guiding me towards this insight.}. This only adds to the complexity of the predictive signal we aim to extract from past conflict patterns. As such, capturing this signal effectively using manually engineered features may simply prove an insurmountable task.\par

The solution I propose here is a bespoke deep learning architecture explicitly designed to automatically learn powerful spatiotemporal features optimized for conflict forecasting. This endeavor also provides an opportune moment to develop and integrate several additional innovative properties warranted by prominent researchers from the field of conflict forecasting \citep{hegre2017introduction, hegre2021viewsappA2020}. 

First, the architecture is engineered to estimate and forecast the patterns of multiple types of violence simultaneously. This design is motivated by findings from previous studies, showing that locations with a high predicted probability of one type of violence also demonstrate an elevated risk for other types of violence \citep{hegre2021viewsappA2020}. Recognizing considerable room for improvement in modeling this interplay, past studies have emphasized the importance of prioritizing and integrating the forecasting of these outcomes into a unified framework \citep{hegre2021viewsappA2020}.\par 

Second, the architecture is also capable of estimating both the probability and magnitude of each type of violence simultaneously. While many existing systems predominantly focus on forecasting the probability of conflict, it is at least as important to consider the expected magnitude of future conflicts. Indeed, identifying areas with a low but genuine risk of extreme violence may be more critical than identifying areas with near-certain but low levels of predicted violence \citep{hegre2017introduction}. Thus, by assessing both the probability and potential magnitude, the architecture provides a more comprehensive understanding of future risks.\par 

Third, during inference, the architecture allows us to draw an arbitrary number of samples from approximate predictive posterior distributions. In other words, for each spatiotemporal unit and target, it forecasts multiple plausible scenarios -- a departure from conventional methods that typically offer singular predictions (a point estimate) for each forecasted unit. This capability enables us to quantify uncertainties surrounding the forecasts, making it a relevant addition to any forecasting framework. It is, however, particularly warranted given conflict forecasting, due to the importance of tail-end outcomes such as rare but devastating events \citep{hegre2017introduction, hegre2021views2020}. In short, the architecture allows decision-makers, practitioners, and researchers to extract valuable insights regarding the credibility of individual predictions and nuanced warning signals about potentially high-impact events.\par

These properties render the architecture highly suitable for conflict forecasting while also showcasing its flexibility. Indeed, the proposed architecture could readily be adapted to accommodate any forecasting task where spatiotemporal patterns are expected to carry relevant predictive signals.\par

%% file: sections/design.tex
I will now introduce the data and its structure, as the latter plays a crucial role in the network architecture. Following that, I will present the network architecture prober, and subsequently, provide a brief overview of the training procedure. First, however, I will offer a high-level intuition.\par

\subsection{The intuition}

Imagine you are watching a movie, and you want to guess what will happen in the next picture frame. You pause the movies and look at the current frame while also taking into account everything that has happened so far. This is a reasonable task as the next frame is bound to be connected to the events unfolding in the previous frames and the overarching plot of the movie. As such, you will likely do quite well in this small prediction task -- even if you are asked to forecast several frames into the \enquote{future}.\par

Now, let us take this analogy a step further and imagine a stack of monochrome satellite images, each measuring $720 \times 360$ pixels. These images depict the globe at discrete monthly intervals, and the pixel values represent the sum of fatalities for specific locations during the corresponding months. Say the stack consists of $120$ such monthly images, sorted from the first observed month to the last. Effectively, this stack of images depicts the history of organized violence across ten years ($10\times12=120$). A toy example of such a stack covering Africa and the Middle East is shown in \autoref{fig:zstack}.\par

\import{figures_tex}{zstack}


If you were to look through such a stack of images from the first to the last, you would find noisy and complex yet discernible spatiotemporal patterns such as clusters, trends, and sporadicity. Some patterns will generalize across the globe, while others will vary across different locations. By examining the first $100$ images and tracing these patterns through space and time, you would likely be able to form a rather qualified guess regarding how the $101^{st}$, $102^{nd}$, and perhaps even the $120^{th}$ image might look. Your predictions would not be perfect, and they would likely get worse as you try to predict further into the future -- but they would be far better than uninformed guessing.\par

My goal here is simple: to create a \enquote{machine} highly specialized for this kind of task. I.e., a machine that can process each \enquote{image}, or \emph{grid}, in temporal sequence and learn highly intricate spatiotemporal patterns directly from the data. Importantly, it must be able to learn patterns that generalize broadly, throughout time and space, while also remembering the specific history of each \enquote{pixel}, or \emph{grid-cell}. When the machine reaches the last observed grid, it should be capable of using this information to generate qualified estimates about the cell-wise patterns in the next, yet unobserved, grid.\par

\subsection{Data, preprocessing, and partitioning}

As this effort exclusively focuses on forecasting future conflict patterns using historical conflict data, I will rely solely on a single data source. \par

Specifically, data on conflict fatality from the Uppsala Conflict Data Program Georeferenced Event Dataset (UCDP GED) \citep{UCDP_2017}, I access data through the VIEWSER API \citep{VIEWSER6}, ensuring streamlined retrieval and direct comparisons with the Violence and Impact Early Warning System (VIEWS) \citep{hegre2019views, hegre2021views2020}. The fatalities are categorized into \emph{state-based} ($sb$), \emph{non-state} ($ns$), and \emph{one-sided categories} ($os$), and aggregated at the monthly PRIO grid cell level ($pgm$) -- with each cell measuring roughly $0.5\times0.5$ decimal degree (roughly $55km\times55km$ at the equator) \citep{Tollefsen_2012}. Currently, VIEWSER covers Africa and the Middle East with a $180\times180$ grid \footnote{Testing with a manually compiled global grid of $360\times720$ cells yielded no substantial changes to the conclusions given below.}.\par

Following \cite{hegre2021views2020}, I partition the data into a calibration partition for experimentation and hyperparameter tuning, spanning from January $1990$ to December $2015$ ($312$ months), and a validation partition for the final evaluation, spanning from January $1990$ to December $2018$ ($348$ months). In both partitions, the last $36$ months are reserved as a hold-out test set and not used for training. As such, each partition contains two subsets: A training set and a test set. The specific partitioning details can be found in Table 1 (see \autoref{tab:table1}).\par

\import{tables}{partition}\par

To conform with established practices in conflict forecasting, the fatality counts are logarithmically transformed\footnote{The logarithmic transformation reduces the impact of extreme outliers and acknowledges that a change from 0 to 100 fatalities might be a more substantial signal than a change from 1000 to 1100 fatalities. Furthermore, machine learning algorithms generally perform better when working with data that is within condensed value ranges \citep[114]{burkov2020machine}. The addition of 1 prevents the logarithm from yielding negative infinity in the absence of any fatalities.} to capture \emph{conflict magnitude} \citep{hegre2019views, hegre2021views2020}. Equation \eqref{eq:targets} defines the conflict magnitude ($cm$) for each of the three types of violence ($sb$, $ns$, $os$) at the $pgm$ level, with months ($t$) and grid-cells ($i$). \par

\[
cm^{sb}_{it} = ln(fatalities^{sb}_{it}+1),\quad
cm^{ns}_{it} = ln(fatalities^{ns}_{it}+1),\quad
cm^{os}_{it} = ln(fatalities^{os}_{it}+1)\quad \tag{1}\label{eq:targets}
\]

For all tasks performed by the network, $cm^{sb}_{it}$, $cm^{ns}_{it}$, and $cm^{os}_{it}$ are the only external features used. For the regression task, the targets will simply be future levels $cm^{sb}_{it}$, $cm^{ns}_{it}$, and $cm^{os}_{it}$, respectively. For the classification task, I create binary targets ($\in \{0,1\}$) by transforming each $cm$ into conflict presence $cp$ using a binary transformation as shown in equation \eqref{eq:bi_targets}. Note that these binary transformations are only used as classification targets and not input features. \par

\[
cp^{sb}_{it} = \begin{cases} 1 & \text{if } cm^{sb}_{it} > \text{0}, \\ 0 & \text{if } cm^{sb}_{it} = \text{0}. \end{cases},\quad 
cp^{ns}_{it} = \begin{cases} 1 & \text{if } cm^{ns}_{it}  > \text{0}, \\ 0 & \text{if } cm^{ns}_{it}  = \text{0}. \end{cases},\quad 
cp^{os}_{it} = \begin{cases} 1 & \text{if } cm^{os}_{it} > \text{0}, \\ 0 & \text{if } cm^{os}_{it} = \text{0}. \end{cases}\quad \tag{2}\label{eq:bi_targets}
\]

Thus, I will use three input features: $cm^{sb}_{it}$, $cm^{ns}_{it}$, and $cm^{os}_{it}$, with six corresponding targets: future values of $cm^{sb}_{it}$, $cm^{ns}_{it}$, and $cm^{os}_{it}$, as well as their binary transformations $cp^{sb}_{it}$, $cp^{ns}_{it}$, and $cp^{os}_{it}$. As detailed more thoroughly below, the model will predict future cell ($pgm$) values using past cell ($pgm$) values. The terms \enquote{past} and \enquote{future} are here left intentionally vague, as the spatiotemporal reach of patterns will be estimated, not assumed, leaving notations like $\hat{cm}^{sb}_{it+1} = f(cm^{sb}_{it}, cm^{ns}_{it}, cm^{os}_{it})$ misleading. Denoting the set of all past $pgm$s as $\iota\tau$, a more accurate though still vague expression could be $\hat{cm}^{sb}_{it+1} = f(cm^{sb}_{\iota\tau}, cm^{ns}_{\iota\tau}, cm^{os}_{\iota\tau})$, and similarly for the other five targets. This formulation recognizes that, in principle, the architecture possesses the capacity to consider any pertinent information, from any past $pgm$, that may contribute to predicting the future patterns of violence in any one specific $pgm$. \par

\subsection{From tabular data frame to z-stack volume}

The proposed network architecture draws inspiration from fields that handle unstructured data like images and text, quite different from conventional tabular datasets. Thus, I need to transform the data from a tabular data frame to a z-stack volume. The analogy to an image stack closely mirrors the actual implementation. Consider, for instance, state-based violence ($cm^{sb}$), represented in a $2D$ data frame. Here, rows are individual observations ($pgm$s) and the columns are the features grid id ($i$), month id ($t$), and the magnitude of state-based violence ($cm^{sb}$). I transform this data frame into a $3D$ volume (or tensor), where the width represents longitude, the height represents latitude, and the depth represents time (months). Each unit (voxel) in the volume corresponds to a $3D$ grid cell with dimensions of $0.5 \times 0.5$ decimal degrees $\times$ $1$ month, and the value of the cell is here $cm^{sb}_{it}$. Thus, both the grid id ($i$) and the month id ($t$) are translated into actual positions in the volumen\footnote{With $i$ now representing a unique combination of latitude and longitude coordinates instead of a single unique cell id}. We can also think of this $3D$ volume as a z-stack of monthly $2D$ grids (images), where each grid represents the monthly conflict patterns $cm^{sb}_t$. Given this perspective, each $cm^{sb}_{it}$ is the value of an individual grid cell (pixel) on a specific monthly grid. \par

To incorporate all three types of violence ($sb$, $ns$, and $os$), I add an extra dimension to the volume, enabling each monthly grid to support multiple \emph{channels}. Given three different types of violence, this extension is akin to going from a stack of monochrome images with one channel each to a stack of color images with three channels each (red, green, and blue). Essentially, I'm allowing each $pgm$ to carry more information, similar to how color images provide more information than monochrome images. Importantly, more channels can readily be added to accommodate additional features, but for the sake of parsimony, I here exclusively include the three types of violence. \par

With this volume structure in place, the next step is to design a \enquote{machine} which can traverse the z-stack sequentially to learn intricate spatiotemporal patterns of violence and subsequently extrapolate such patterns into further months. \par

\subsection{Introducing the HydraNet Architecture}

Two factors need highlighting. First, the fundamental ideas behind the proposed architecture extend to spatiotemporal forecasting tasks more broadly. The specific implementation presented here, denoted \emph{HydraNet}, is primed for the task of conflict forecasting using past conflict patterns as predictors. 

Second, the architecture is an advanced neural network that builds upon several very active and ambitious disciplines. Thus, given its complexity, curious readers should consult the online appendix for an in-depth account of its structure. The aim here is to provide a baseline intuition regarding how and why the architecture works. \par

Now, the core idea is to create an architecture that is capable of learning detailed spatiotemporal features optimized for forecasting future conflict automatically. Given the proven effectiveness of simple lags and decay functions as predictive features \citep{hegre2019views, vesco2022united}, I start with the assumption that temporal and spatial distances matter. Beyond this, I want the particularities of any relevant pattern to be learned -- not assumed. To this end, I design the architecture around four key attributes: it is a \emph{deep}, \emph{convolutional} neural network, organized in a recurrent \emph{long short-term memory} structure, that employs \emph{skip connections}. \par

The deep architecture is a key advantage of any neural network: By having successive layers of parameters utilize the outputs of preceding layers, deep architectures gain the capacity to capture progressively intricate patterns, enabling the network to learn complex relationships between input and target automatically \citep{goodfellow2016deep, burkov2020machine}. This eliminates the need for manually engineered features -- such as temporal and spatial lag -- and ensures that the learned features are inherently optimized for the task. \par

The trade-off is that learning these features using traditional \emph{densely connected feedforward} networks requires vast amounts of data -- often hundreds of millions of examples. Yet, even with highly granular units like $pgm$s, conflict events are rare. To address this, the proposed architecture adopts design elements from architectures specialized for learning sequential and spatial features efficiently, namely Long Short-Term Memory neural networks (LSTMs) and Convolutional Neural Networks (CNNs), respectively. \par

HydraNet is constructed as a CNN, meaning it consists of convolutional layers designed to autonomously learn intricate spatial dependencies in $2D$ grids \citep[321]{goodfellow2016deep, brodrick2019uncovering}. Since the objective is to predict cell-wise patterns on an output grid with the same spatial dimensions as the input grid, I constructed the architecture similarly to the \emph{encoder-decoder} networks used for \emph{semantic segmentation} \citep{badrinarayanan2017segnet}.\par

In our case, consider the simplified task of predicting the future monthly $2D$ grid, $cm^{sb}_{t+1}$, using only $cm^{sb}_{t}$. Similar to moving a magnifying glass over a map, convolutional layers slide filters -- small trainable squares of parameters -- across the input grid to identify salient spatial patterns. During training, these parameters are optimized to map the input grid to the target grid. By utilizing \emph{parameter sharing}, i.e., applying the same filters iteratively across the entire input grid, the network can identify recurring patterns across the full spatial context, rather than treating each location (i.e., grid cell) in isolation. This approach greatly enhances the network's capacity to efficiently learn salient spatial patterns. To learn multiple different spatial patterns, each layer of a CNN contains multiple filters, and by consecutively stacking convolutional layers in a deep architecture, the network can automatically learn increasingly complex spatial patterns \citep[322-328]{goodfellow2016deep}.\par

A side-effect of increasing the depth of a CNN is the loss of detailed cell-specific information. For CNNs designed to capture larger visual features in images, this is not an issue. However, given the task of conflict forecasting, where each grid cell represents a substantial geographical area, precise pixel-wise estimation must be prioritized. To solve this issue, I adopt a \emph{U-net} architecture. Designed for preserving detailed spatial information in limited data contexts like medical imagery, this design employs skip connections between convolutional layers to retain cell-wise precision through deep architectures \citep{ronneberger2015u, kohl2018probabilistic}. Indeed, while the subject matter of medical imagery and conflict forecasting differ, the task is rather similar: with limited data, learn complex spatial dependencies mapping an input grid to a target output grid while prioritizing cell-wise precision. \par

Now, training a network that uses a $2D$ grid at month $t$ to predict a subsequent grid at month $t+1$ is a good start. Indeed, training the same network sequentially on monthly input-target pairs, $t$ to $t+1$, then $t+1$ to $t+2$, and so on, constitutes parameter sharing across time without potentially violating the direction of causality \citep[362-372]{goodfellow2016deep}. This is akin to flipping through a stack of monthly maps depicting conflict and learning general patterns connecting the patterns of the current month to those of the proceeding month.\par

What is missing from this design is a way to incorporate specific $pgm$ level history, such as long and short-term trends. Any future state, such as the grid $cm^{sb}_{t+1}$, is likely influenced by more than just the current state $cm^{sb}_{t}$. To address temporal dependencies, common practices involve utilizing lags and decay functions to construct features containing information on past patterns.\par

Deep LSTM networks bypass the need for such manual feature engineering by carrying a \emph{hidden state} and a \emph{cell state} across each step of the sequence. Given forecasting, we can think of the hidden state as short-term memory ($s_t$) and the cell state as long-term memory ($l_t$). At each step in the sequence, an LSTM \emph{cell} uses these memory states to transform the input, creating a representation that incorporates relevant information from preceding steps -- i.e., the \enquote{past}. Simultaneously, it updates both $l_t$ and $s_t$ to $l_{t+1}$ and $s_{t+1}$, using any relevant signals from $cm^{sb}_{t}$ \citep[400]{goodfellow2016deep}. Crucially, the network automatically learns which temporal patterns should be remembered, updated, used, and forgotten, given the specific prediction task at hand \citep[363-366]{goodfellow2016deep}. A single LSTM cell can learn complex and varied long- and short-term signals, but using multiple LSTM cells in parallel extends this capacity markedly. These elements are all part of the LSTM design to allow for effective handling of sequential patterns -- and its success is extensively demonstrated in forecasting time-series data, including predicting violent conflicts \citep{hochreiter1997long, malone2022recurrent, radford2022high}.

Thus, to create a highly effective framework for conflict forecasting, the HydraNet architecture is essentially a deep convolutional U-net modified to fit into an LSTM structure. The network's depth facilitates automatic feature generation, and the convolution layers enable efficient learning of complex spatial patterns. The skip connections of the U-net ensure cell-wise precision is retained through the deep architecture, and the LSTM structure facilitates the efficient learning of temporal features. As these four elements are seamlessly incorporated into the architecture, it is capable of processing a sequence of grids and learning highly intricate spatiotemporal patterns directly from the data. It learns patterns that generalize broadly, throughout time and space, while also retaining and updating salient historical information. Upon reaching a final observed grid, the architecture can utilize the parameters learned and the information retained to produce predictions regarding cell-wise patterns in the subsequent, yet unobserved, grid.\par

Two additional attributes enhance the architecture for the task at hand. First, I want to handle three types of violence -- $cm^{sb}$, $cm^{os}$, and $cm^{ns}$ simultaneously. I also want to solve both classification and regression tasks for each violence type. To achieve this, the network is designed with multiple decoders, optimized as a multitasking network \citep{kendall2018multi}. In essence, it uses a shared encoder for three input features, while generating six distinct outputs through six separate decoders. This design enables the joint optimization of all parameters while still allowing for distinct tasks.\par

Secondly, to quantify some of the uncertainty inherent to conflict forecasting \citep{hegre2017introduction}, I incorporate active \emph{dropout layers} throughout the network architecture \citep{gal2016dropout, kendall2015bayesian}. These layers serve to prevent overfitting, aid feature learning during training, and allow for the generation of non-deterministic predictions during test time. Essentially, this allows me to draw predictions as Monte Carlo samples over the model space, providing a cost-effective approximation of a Bayesian predictive posterior distribution \citep{gal2016dropout}. Thus, capturing some amount of model uncertainty.\par

\import{figures_tex}{hydranet_simple}

\autoref{fig:hydranet} show a simplified schematic of the network architecture -- a more detailed version can be found in the online appendix.\par

\subsection{The training process}

Following \cite{hegre2019views} and \cite{hegre2021views2020}, the training routine consists of two phases: calibration and validation. During calibration, non-learnable hyperparameters -- such as network depth, loss functions, and training epochs -- are fine-tuned using the calibration partition. Specifically, the network is iteratively trained on the calibration training set and evaluated on its hold-out test set until a satisfactory set of hyperparameters is identified. Detailed specifications are provided in the online appendix. In the second phase, the network is trained once, from scratch, on the training set of the validation partition using the calibrated hyperparameters. Final performance is then evaluated using the hold-out test set from the validation partition. \par

I do not train the network on the full spatial extent of the data volume, as this would significantly increase computational cost and risk overfitting -- or, conversely, model collapse due to the predominance of zero-valued cells. Instead, training is performed on smaller spatial subsets. For example, while the full training tensor from the calibration partition spans $180 \times 180 \times 276$ grid cells, the model is iteratively trained on spatial patches of $32 \times 32 \times 276$.

To ensure that these patches contain informative signal, I adopt a Predefined Curriculum Learning strategy \citep{bengio2009curriculum, wang2021survey, soviany2022curriculum}. Early in training, sampling is biased toward regions with observed fatalities. As training progresses, the sampling distribution is gradually widened to include more diverse spatial subsets. The specific sampling procedure is described in the online appendix, and an illustration of the sampled patches is shown in \autoref{fig:patches}.

\import{figures_tex}{patches}

Despite the targeted sampling strategy, the dataset remains highly imbalanced -- most locations record no fatalities during most time steps. To mitigate this, I employ specialized loss functions tailored to the nature of each task: a shrinkage loss for the three regression tasks \citep{lu2018deep}, and a focal loss for the three classification tasks \citep{lin2017focal}. Both losses are designed to prioritize rare events and hard-to-predict cases. Following \cite{kendall2018multi}, the six individual losses are combined into a unified multi-task loss function. This approach allows the network to jointly optimize all targets while explicitly addressing the underlying data imbalance. \par

During out-of-sample evaluation -- both in the calibration and validation phases -- the network is exposed to the full data volume of the relevant partition, including the hold-out test set months. It sequentially processes this volume from the first to the last month of the training set, leveraging its learned parameters to identify relevant spatiotemporal patterns and update its internal representations -- i.e., the hidden and cell states, corresponding to short- and long-term memory, respectively.

Upon reaching the end of the training period, the model transitions into testing by switching from observed inputs to autoregressive predictions -- feeding its own past outputs back as inputs. This setup enables the architecture to generate forecasts of arbitrary lengths, although predictive performance naturally deteriorates with increasing temporal distance. In line with \cite{hegre2021views2020}, I set the forecast horizon to 36 months.

Lastly, by keeping dropout active during testing, I introduce controlled stochasticity into the network’s output. This technique -- known as Monte Carlo Dropout \citep{gal2016dropout} -- approximates sampling from a diverse ensemble of models, thereby enabling the quantification of model uncertainty. In practical terms, each forward pass through the network yields a different prediction, reflecting variability in the model space. This allows me to generate 36-month out-of-sample forecasts as an approximate Bayesian predictive posterior distribution. For each forecasted value, I draw 128 samples, with their mean serving as the point estimate for downstream analysis and evaluation.

%% file: figures_tex/zstack.tex
\begin{figure*}[!ht]
\centering

\includegraphics[width=12cm]{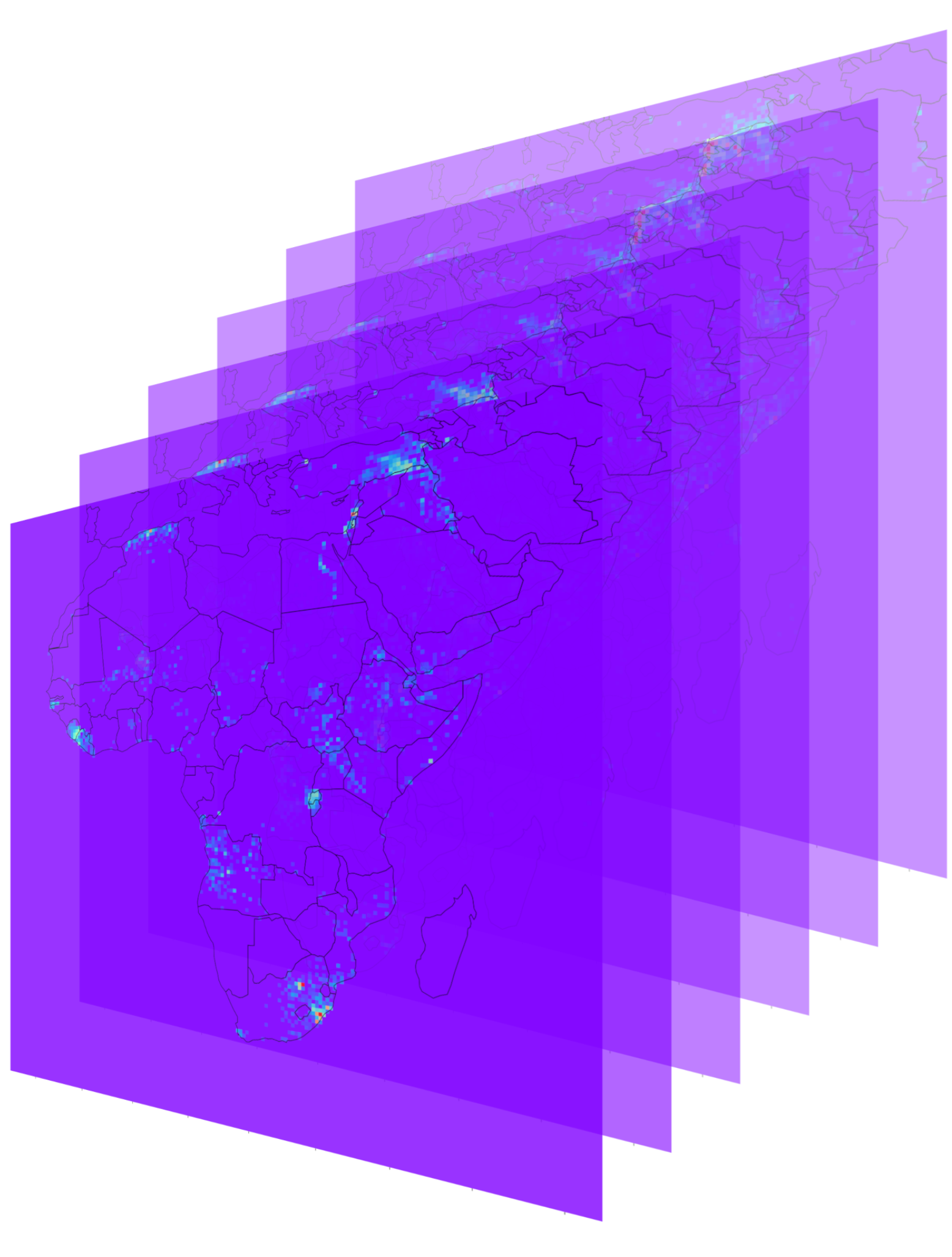}
\caption{Toy example of a z-stack visualized as a series of 2D grids covering Africa. Each grid represents a unique time step, and each \enquote{pixel} corresponds to a PRIO grid cell. To enhance visual clarity, borders, and coastlines are overlaid, conflict occurrences are aggregated, and the values are condensed.}\label{fig:zstack}
\end{figure*}

%% file: tables/partition.tex
\begin{table}[ht!]
  \begin{center}
    \begin{tabular}{p{0.2\linewidth} P{0.16\linewidth} P{0.16\linewidth} P{0.16\linewidth} P{0.16\linewidth}}
      \toprule
      \textbf{Partitions}& \thead{Start\\training} & \thead{End\\training} & \thead{Start\\testing} & \thead{End\\testing} \\
      \midrule
      Calibrating              & 01/01/1990       & 31/12/2012    & 01/01/2013      &  31/12/2015 \\
      Validating                     & 01/01/1990     & 31/12/2015   & 01/01/2016   & 31/12/2018 \\
      Forecasting                 & 01/01/1990     & 31/12/2018 & - & - \\ 
      \bottomrule
    \end{tabular}
    \caption{A 36-month test set within each partition is not used for training. Notably, the test set for calibration concludes before the test set for validation initiates. This is all done to ensure a valid and honest evaluation of the model's performance.
    }\label{tab:table1}
  \end{center}
\end{table}

%% file: figures_tex/hydranet_simple.tex
\begin{figure*}[!ht]
\centering

\includegraphics[width=16cm]{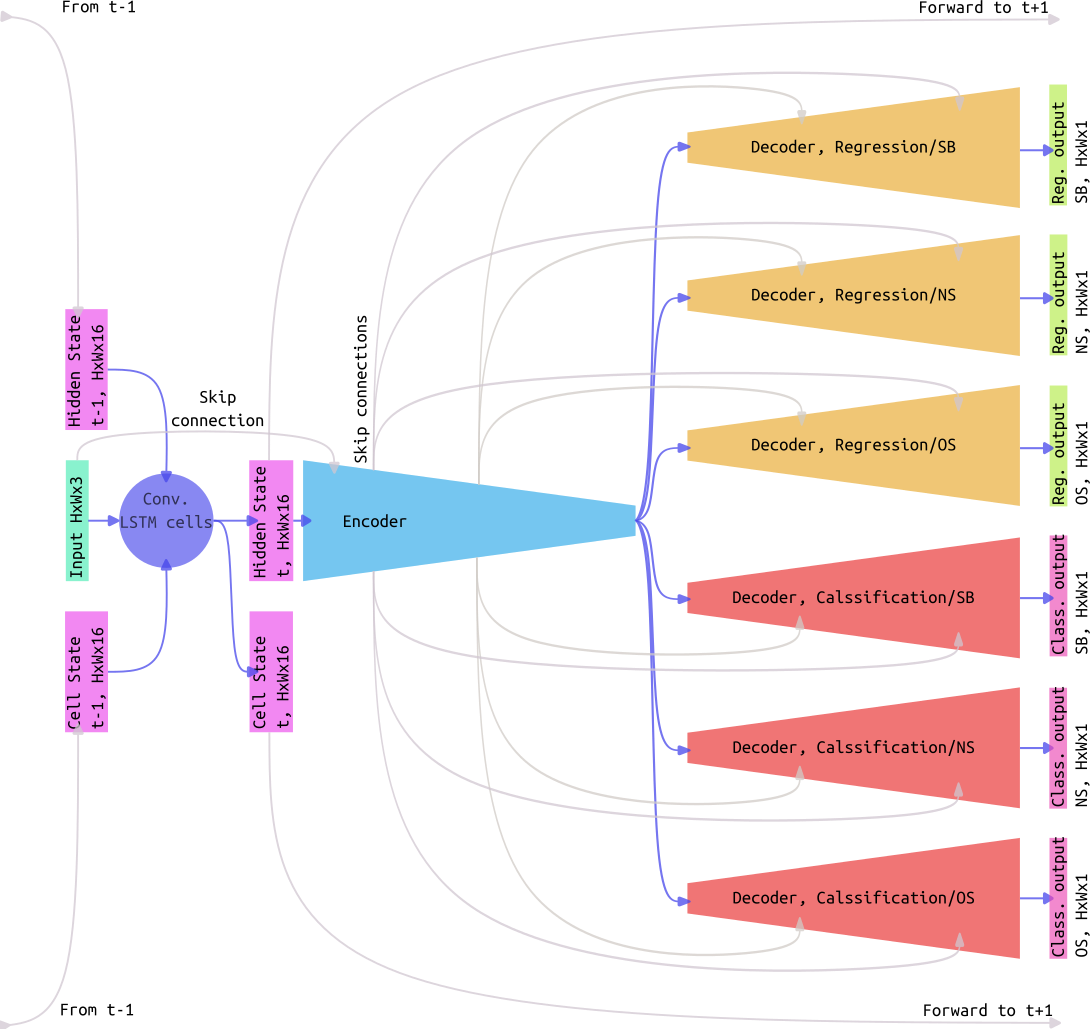}
\caption{Simplified schematic of HydraNet architecture featuring a deep convolutional encoder/decoder structure with skip connections, organized in a recurrent LSTM structure. The shared encoder handles all three violence types (state-based, one-sided, and non-state-based), and dedicated decoders estimate both the probability and magnitude of violence for each type. Thus, the input grid/tensor has three channels, while each of the six output grids/tensors has one. The network parameters are collectively optimized through a multi-task loss function.}\label{fig:hydranet}
\end{figure*}

%% file: figures_tex/patches.tex
\begin{figure*}[!ht]
\centering

\includegraphics[width=14cm]{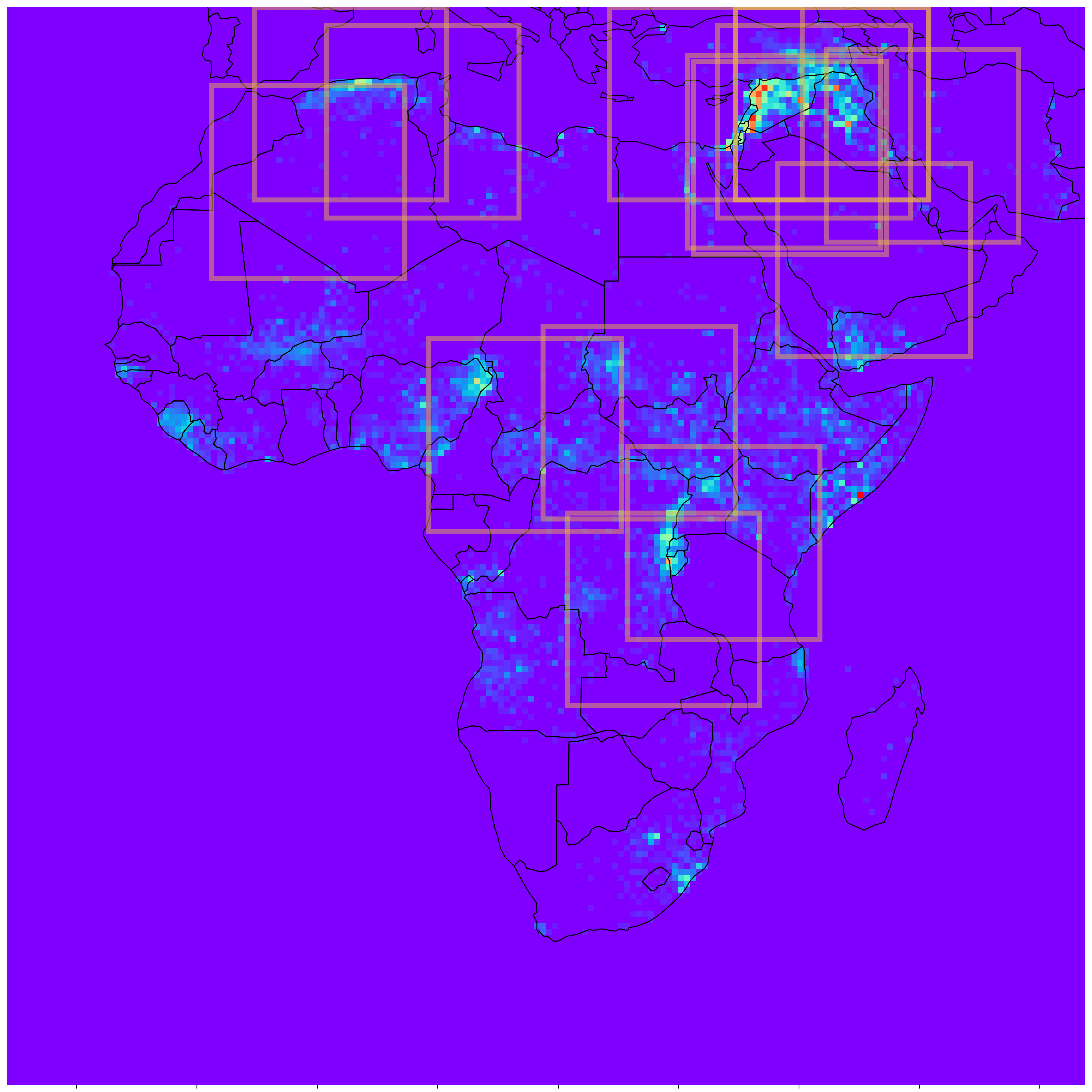}
\caption{A toy example of the patch sampling process. The orange squares represent $15$ sampled patches for $32\times32$ grid cells. Note that each sampled patch stretches through all months in the training set. For instance, sampling a patch of $32\times32$ grid cells from a $180\times180\times312$ z-stack would result in a $32\times32\times312$ tensor, with the last dimension being months. To enhance visual clarity, borders, and coastlines are overlaid, conflict occurrences are aggregated, and the magnitude range is condensed.}\label{fig:patches}
\end{figure*}



%% file: sections/evaluation.tex
With model training complete and samples generated from the predictive posterior, I now turn to assessing the predictive performance of the proposed architecture. Unless otherwise noted, all results reported below are based exclusively on out-of-sample performance -- specifically, those derived from the test set of the validation partition. This ensures that evaluation is conducted on data that the model was neither directly trained on nor indirectly exposed to during hyperparameter tuning.

The next subsection briefly introduces the metrics used to evaluate HydraNet’s performance. A detailed technical exposition of these metrics lies beyond the scope of this article, but interested readers can find additional information in the online appendix. Following this overview, I present the results and compare HydraNet’s performance to that of the current state-of-the-art framework, VIEWS. \par

\subsection{Evaluation Metrics}



To assess HydraNet’s predictive performance, I focus on evaluation based on point estimates,\footnote{While the architecture produces full posterior distributions via Monte Carlo Dropout sampling, the focus here is on evaluating point estimates. This decision was made to maintain comparability with \cite{hegre2021views2020}, which also reports point-based metrics. Nonetheless, I recognize the importance of distributional evaluation (e.g., using $CRPS$) and plan to include such analysis in a future revision of this working paper.} in line with the VIEWS framework \citep{hegre2019views, hegre2021views2020}, which also reports point-based metrics. Specifically, I compute the mean of 128 posterior samples for each spatiotemporal unit and evaluate performance using standard regression and classification metrics. For regression tasks, I report the Mean Squared Error ($MSE$). For classification tasks, I employ the Area Under the Receiver Operating Characteristic Curve ($AUC$), Average Precision ($AP$), and the Brier Score \citep{hegre2019views}. Historically, $AP$ has served as the primary evaluation criterion in this domain \citep{hegre2019views}, and to ensure comparability with prior work, I adopt this convention. Accordingly, the network presented here is optimized with a principal focus on maximizing $AP$.

As a reminder, both $AUC$ and $AP$ scores range from 0 to 1, with values closer to 1 indicating better classification performance. An $AUC$ of 0.5 reflects performance equivalent to random guessing, while interpreting $AP$ requires contextual grounding \citep[350–351]{su2015relationship}. Specifically, the baseline for random $AP$ performance corresponds roughly to the prevalence of positive events in the dataset \citep[132]{bestgen2015AP}, which, in this case, is approximately 0.005. For the regression metrics -- $MSE$ and the Brier Score -- optimal performance is indicated by values approaching zero.

\subsection{Results}

To evaluate HydraNet under the most stringent conditions, I benchmark its performance against the best-performing models and ensembles reported in \cite{hegre2021views2020} and detailed in Appendix D of \cite{hegre2021viewsappD2020}. For example, the $AP$ score for state-based violence is taken from the ensemble identified by \cite{hegre2021views2020} as achieving the highest performance across all tested configurations. This benchmarking strategy is applied consistently across all VIEWS-related metrics.

In contrast to the ensemble-based approach used by VIEWS, the results for HydraNet are derived from a single model trained with a fixed set of hyperparameters\footnote{Training was conducted over approximately two hours on an NVIDIA Tesla V100-PCIE-16GB Tensor Core GPU}. To account for stochastic variation in neural network training, I trained 12 independent HydraNet instances and selected the one whose performance was closest to the sample mean. All reported results are based on this representative model\footnote{After the implementation of Curriculum Learning, the variability across trained models was negligible}. This approach is intended to provide a realistic estimate of the architecture’s typical performance. Summary statistics for all metrics -- averaged across out-of-sample months -- are presented in \autoref{tab:mean_results_no_me}.

\import{tables}{mean_results_no_me}

Notably, both HydraNet and the VIEWS framework substantially outperform the no-change baseline across all tasks -- with the partial exception of the Brier score, where the baseline remains competitive due to its conservative forecasting.

Leveraging a large feature set -- including conflict and protest history, geographic variables, economic indicators, and political institutions -- the VIEWS framework achieves strong performance across all three violence categories \citep{hegre2019views, hegre2021viewsappC2020}. For state-based violence, it reports an $AUC$ of 0.921, an $AP$ of 0.272, and a Brier score of 0.0046. For non-state-based violence, the corresponding scores are 0.894 ($AUC$), 0.047 ($AP$), and 0.0020 (Brier score). For one-sided violence, VIEWS achieves an AUC of 0.900, an AP of 0.138, and a Brier score of 0.003 \citep{hegre2021viewsappD2020}.

In contrast, HydraNet relies solely on spatiotemporal patterns of past conflict -- excluding structural covariates such as political or economic indicators -- yet outperforms VIEWS across all three violence types on $AP$, the primary evaluation metric for both frameworks. Specifically, HydraNet achieves an $AP$ of 0.304 for state-based violence, 0.134 for non-state-based violence, and 0.162 for one-sided violence.

Beyond classification, HydraNet also generates continuous-valued regression estimates of conflict intensity and produces full posterior distributions over predicted outcomes via Monte Carlo Dropout sampling. Neither of these capabilities is supported in the 2020 VIEWS specification\footnote{Note that ongoing research within the VIEWS project is actively addressing these limitations \citep{hegre2021views2020}}. Together, these results underscore HydraNet’s capacity to extract meaningful predictive signals from temporal and spatial conflict dynamics -- despite relying on a dramatically simplified input space.

To ensure direct comparability with the VIEWS 2020 results \citep{hegre2021views2020}, all evaluation scores for HydraNet in \autoref{tab:mean_results_no_me} exclude the Middle East region. This aligns with the original VIEWS validation setup, which similarly omitted the region. For completeness, results from a HydraNet specification that includes the Middle East are reported in the appendix (\autoref{tab:mean_results_me}). While these are not used for benchmarking against VIEWS, they show that HydraNet continues to outperform a no-change baseline across tasks and metrics. \par

The appendix provides disaggregated monthly evaluation metrics for each violence type -- state-based, non-state, and one-sided -- with and without the Middle East. These plots illustrate how forecast performance evolves over time and varies by task. State-based violence shows a clear and steady degradation trend, non-state violence is comparatively more stable but declines in the latter half, and one-sided violence exhibits persistent volatility -- underscoring the distinct challenges each category presents. \par

Appendix plots for state-based violence (\autoref{fig:sb_eval}) reveal consistent temporal degradation across all metrics. Mean Squared Error and Brier Score rise sharply after the first year, while $AP$ drops from approximately 0.6 to below 0.25 by the end of the forecast window. $AUC$ remains relatively high but shows a gradual loss of discriminative strength. This pattern reflects the expected challenges of long-horizon forecasting in partially stochastic systems. \par

Appendix figures for non-state violence (\autoref{fig:ns_eval}) display more gradual but still noticeable performance deterioration. $AP$ falls from around 0.3 to below 0.15 after month 12, and $AUC$ begins to fluctuate more significantly in the second half. Regression metrics ($MSE$, Brier) remain relatively stable early on but show moderate increases toward the end of the horizon -- suggesting reduced predictive confidence over time. \par

As shown in \autoref{fig:os_eval}, one-sided violence presents the most erratic performance across the forecast window. Metrics fluctuate substantially month to month, and no clear degradation trend emerges -- suggesting that the model struggles to learn stable patterns in this particularly sparse and irregular data. \par



These challenges are not unique to HydraNet. Similar forecasting degradation patterns -- especially the early success and later collapse in non-state violence -- were observed in the VIEWS 2020 ensemble models \citep{hegre2021views2020}. The observed differences in performance across violence types, therefore, likely reflect structural limitations in the available data, rather than any single modeling approach. These findings highlight an intuitive trade-off between forecast length and model reliability. While early months benefit from strong memory and contextual anchoring, longer horizons suffer from accumulated uncertainty -- some of which remains invisible in the current posterior structure. This underscores the potential for architectural enhancements that explicitly model temporal distance or uncertainty inflation across recursive steps. \par

To simultaneously assess temporal and spatial dynamics, an indicative time-lapse video illustrating the evolution of conflict throughout the test period is available at \href{https://github.com/views-platform/views-hydranet}{https://github.com/views-platform/views-hydranet}.\footnote{\url{https://github.com/views-platform/views-hydranet}} Several aspects warrant attention. \par

First, while HydraNet clearly outperforms the no-change baseline, its own forecasts display a degree of inertia. This behavior appears to stem from two design choices. The first is the decision to freeze the cell state -- the long-term memory -- during the forecasting phase. As a result, although predicted values are recursively fed back as inputs, the persistent cell state acts as a stabilizing prior, anchoring predictions to historical trends. In frequentist terms, this functions as a form of regularization, helping to prevent volatile feedback loops as the forecast horizon lengthens. While this design improves robustness, it may also limit the model's flexibility -- suggesting that a more balanced memory update mechanism could further enhance performance. \par

Second, conflict dynamics themselves are known to exhibit considerable inertia. As shown in \citep{hegre2022lessons}, this relative stability explains why simple history-based models -- such as the no-change baseline -- often perform surprisingly well. Yet, as demonstrated above, HydraNet consistently outperforms such models (as does the VIEWS framework \citep{hegre2019views}). The key insight here is that while conflict is inert, it is not static -- and it is precisely this nuanced temporal structure that HydraNet appears able to capture effectively. \par

%% file: tables/mean_results_no_me.tex
\begin{table}[ht!]
  \begin{center}
    \begin{tabular}{p{0.20\linewidth} p{0.24\linewidth}  p{0.14\linewidth} p{0.14\linewidth} p{0.14\linewidth} }
    \toprule
        \textbf{Violence type}  & \textbf{Task/Metrics}                 & \textbf{No-Change}        & \textbf{VIEWS}                & \textbf{Hydra Net} \\
        
        \midrule
        State-Based             
                                &  MSE                                  & 0.01                      & NA                            & 0.004 \\
                                & \textbf{AP}               & \textbf{0.13} &  \textbf{0.272}   & \textbf{0.304}   \\
                                &  AUC                                  & 0.62                      & 0.921                         & 0.907 \\
                                &  Brier                                & 0.007                     & 0.005                        & 0.009 \\
        \midrule
        Non-State-Based         
                                &  MSE                                 & 0.01                       & NA                            & 0.002 \\
                                &   \textbf{AP}           &  \textbf{0.04}  &  \textbf{0.047}   & \textbf{0.134}        \\
                                &  AUC                                 & 0.53                       & 0.894                         & 0.929        \\
                                &  Brier                               & 0.042                      & 0.002                        & 0.004          \\
        \midrule
        One-sided               
                                &  MSE                                 & 0.01                       & NA                            & 0.003      \\
                                & \textbf{AP}              &  \textbf{0.06} & \textbf{0.138}    & \textbf{0.162}      \\
                                &  AUC                                 & 0.61                       & 0.900                         & 0.884           \\
                                &  Brier                               & 0.006                      & 0.003                        & 0.007        \\
        \bottomrule
    \end{tabular}
   \caption{Evaluation scores for HydraNet, the VIEWS 2020 model \citep{hegre2021views2020}, and a no-change benchmark, across three types of violence: state-based, non-state, and one-sided. Metrics include Mean Squared Error ($MSE$), Average Precision ($AP$), Area Under the Curve ($AUC$), and Brier score, all calculated over the 36-month test set in the validation partition. The Middle East (ME) region is excluded from all models to ensure direct comparability with the \cite{hegre2021views2020} results.}\label{tab:mean_results_no_me}
  \end{center}
\end{table}

%% file: sections/conclusion.tex
This paper introduces HydraNet, a novel probabilistic spatiotemporal forecasting architecture tailored for conflict forecasting. By organizing convolutional layers within a recurrent U-Net structure, the model captures complex spatial patterns and evolving temporal dynamics from raw input data -- without the need for handcrafted features. The resulting framework is capable of forecasting not just the likelihood but also the intensity of future conflict, across three distinct types of political violence.

Importantly, HydraNet produces a predictive posterior distribution over all forecasts using Monte Carlo Dropout sampling, enabling downstream users to quantify uncertainty alongside point estimates. This probabilistic framing is essential in contexts where decision-making often hinges on the range of plausible futures rather than a single deterministic output.

Despite relying on only three input features -- past observations of conflict -- HydraNet achieves state-of-the-art performance across all tasks, outperforming both simple baselines and the best-performing ensembles reported in \cite{hegre2021views2020} on the field’s primary evaluation metric, Average Precision. These gains are achieved while maintaining transparency, modularity, and computational tractability, with the architecture trained end-to-end in under two hours.

Together, these findings underscore HydraNet's capacity to learn meaningful, data-driven representations of conflict dynamics and to forecast complex spatiotemporal outcomes with both accuracy and calibrated uncertainty. Future iterations may further improve performance by incorporating horizon-aware mechanisms, adaptive memory updates, or exogenous structural covariates. Nonetheless, this work provides a strong foundation for principled, probabilistic forecasting of political violence using deep neural networks.

%% file: sections/replication.tex
All code required to replicate the results presented in this paper is written in Python and is available across four public repositories maintained by the VIEWS platform: \par

\begin{itemize}
  \item \textbf{HydraNet} -- core model implementation and training routines: \url{https://github.com/views-platform/views-hydranet}
  \item \textbf{views-models} -- standardized interface and configuration layer for all forecasting models: \url{https://github.com/views-platform/views-models}
  \item \textbf{views-pipeline-core} -- end-to-end forecasting pipeline including data ingestion, preprocessing, and orchestration: \url{https://github.com/views-platform/views-pipeline-core}
  \item \textbf{views-evaluation} -- evaluation workflows used to generate all reported results: \url{https://github.com/views-platform/views-evaluation}
\end{itemize}

A high-level overview of the VIEWS code ecosystem and design structure is available at: \url{https://github.com/views-platform}.

Processed data is accessed via the VIEWSER API -- the official data interface of the VIEWS project. A static snapshot of the data used in this study will be made available at \url{https://www.prio.org/data} and here \url{https://viewsforecasting.org/publications/supplementary-material/} upon publication. Step-by-step replication instructions are included in the HydraNet repository. \par

%% file: sections/acknowledgements.tex
This paper draws on research originally conducted as part of the author’s Ph.D. dissertation at the University of Copenhagen. The author is grateful for the guidance and support of supervisors Lene Hansen and Jacob Gerner Hariri, and thanks the dissertation committee  --  Frederik Hjorth, Lisa Hultman, and Nils Weidmann  --  for valuable feedback.

Subsequent development and revision of this work have been supported by ongoing research at the Peace Research Institute Oslo (PRIO). The author is particularly thankful for discussions with colleagues and collaborators at PRIO, the Center for Social Data Science (SODAS) at the University of Copenhagen, and beyond.

This research has received support from multiple projects: the dissertation was developed with funding from Bodies as Battleground: Gender Images and International Security (Independent Research Fund Denmark, PI: Lene Hansen); the current revision is supported by Societies at Risk: The Impact of Armed Conflict on Human Development (Riksbankens Jubileumsfond, PI: Håvard Hegre); and additional work has been carried out under ANTICIPATE: Anticipating the Impact of Armed Conflict on Human Development (European Research Council, PI: Håvard Hegre).

%% file: sections/appendix.tex
An extended online appendix, including additional plots, model diagnostics, and detailed implementation notes, is available at: \href{https://github.com/views-platform/views-hydranet}{\texttt{github.com/views-platform/views-hydranet}}. 

For convenience, selected key tables and figures are also included below.

\import{tables}{mean_results_me}

\begin{figure}[htb!]
    \begin{minipage}[b]{1.0\linewidth}
      \centering
      \centerline{\includegraphics[width=16cm]{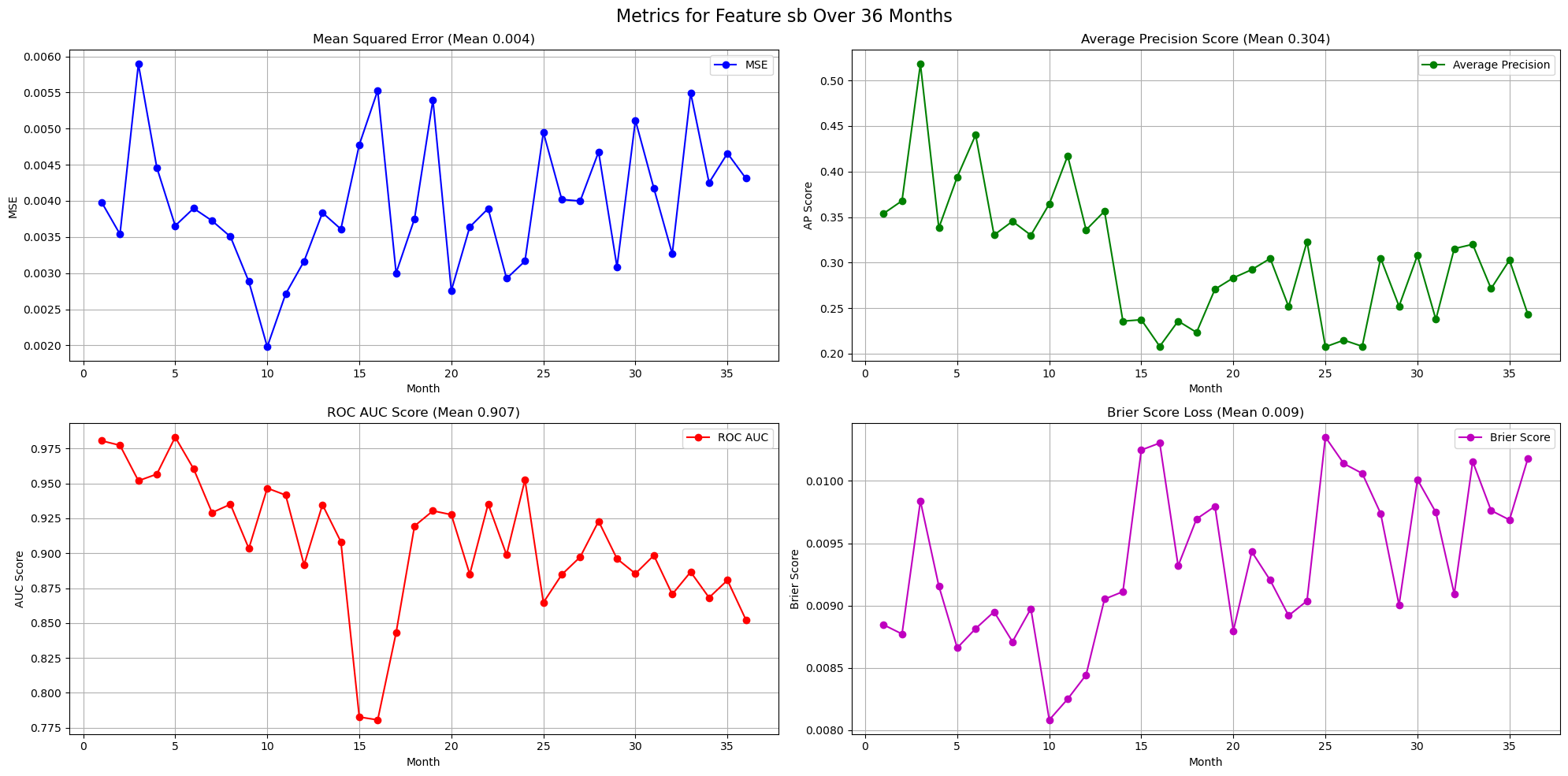}}
    \end{minipage}
    \caption{MSE, AP, AUC, and Brier scores for state-based violence over the 36 hold-out test months in the validation partition. This plot omits the Middle East (ME) to be comparable with \cite{hegre2021views2020} and \cite{hegre2019views}. Point predictions are based on the monthly mean of the sampled distribution.
}\label{fig:sb_no_me_eval}
\end{figure}

\begin{figure}[htb!]
    \begin{minipage}[b]{1.0\linewidth}
      \centering
      \centerline{\includegraphics[width=16cm]{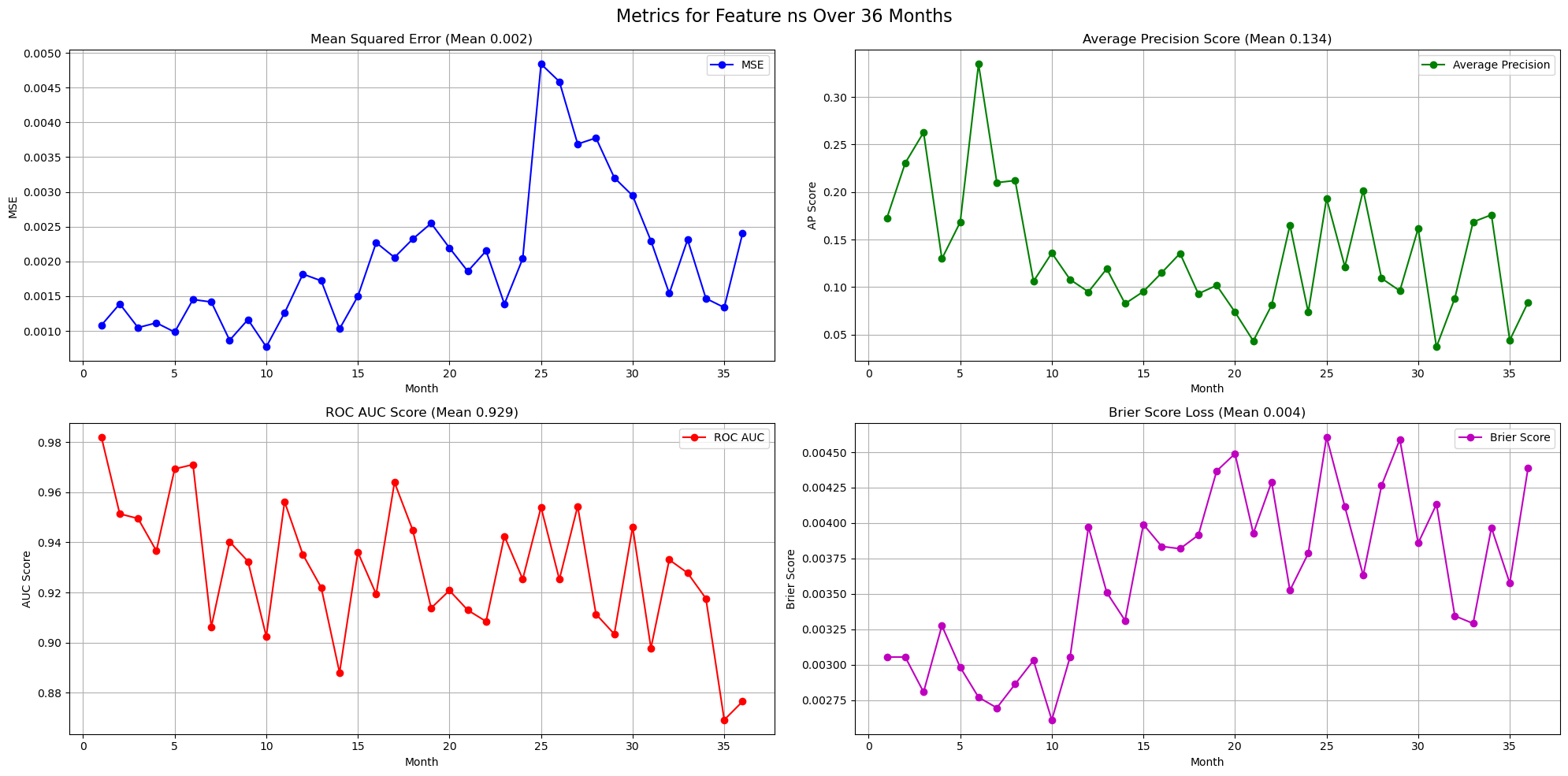}}
    \end{minipage}
    \caption{MSE, AP, AUC, and Brier scores for non-state violence over the 36 hold-out test months in the validation partition. This plot omits the Middle East (ME) to maintain direct comparability with \cite{hegre2021views2020} and \cite{hegre2019views}. Monthly means of the predictive distribution are used as point estimates.
}\label{fig:ns_no_me_eval}
\end{figure}

\begin{figure}[htb!]
    \begin{minipage}[b]{1.0\linewidth}
      \centering
      \centerline{\includegraphics[width=16cm]{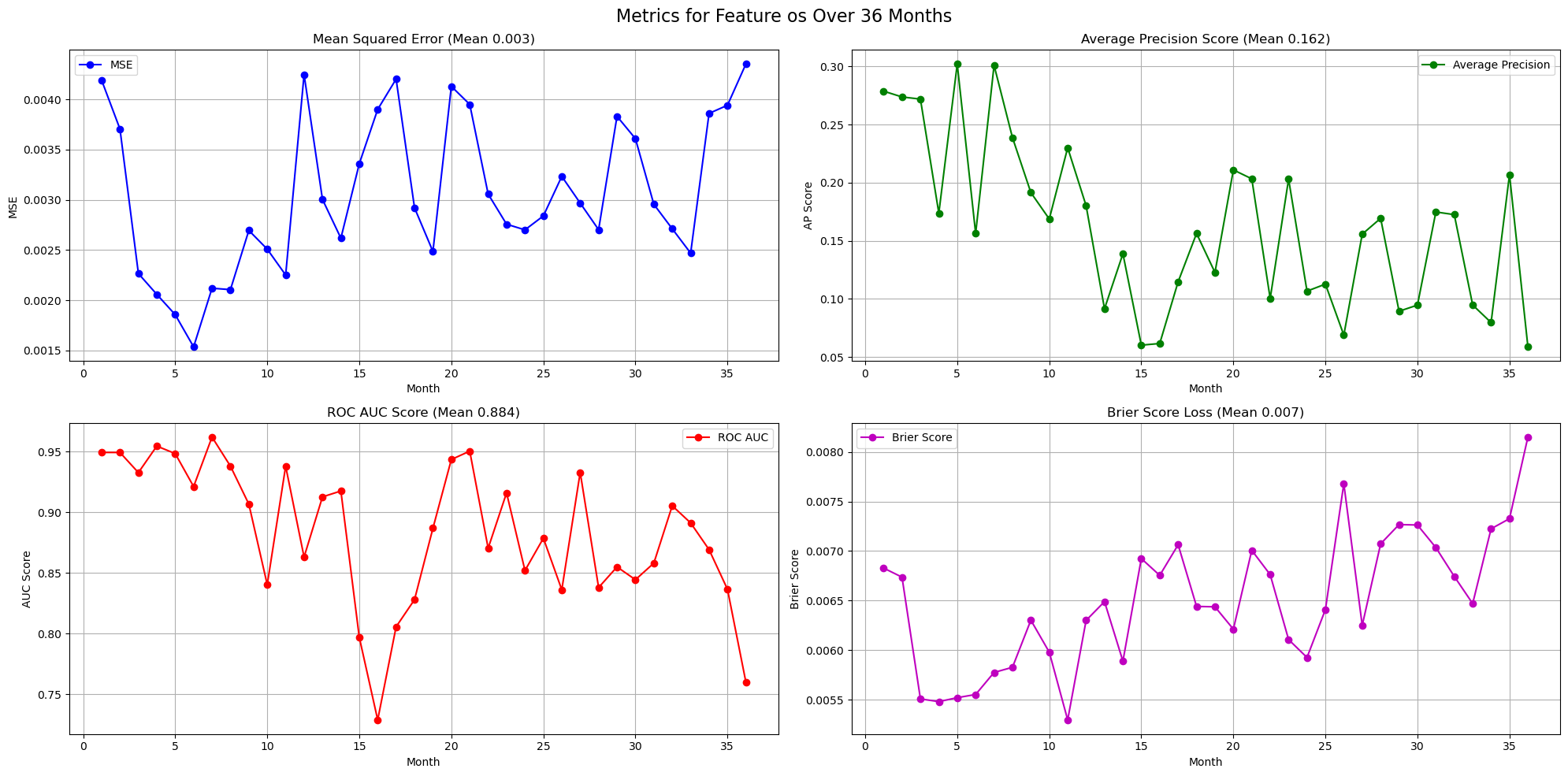}}
    \end{minipage}
    \caption{MSE, AP, AUC, and Brier scores for one-sided violence across 36 validation months in the test partition. The Middle East (ME) is excluded to enable comparison with earlier studies \cite{hegre2021views2020, hegre2019views}. All point estimates reflect the monthly mean of the sampled forecast distribution.
}\label{fig:os_no_me_eval}
\end{figure}

\begin{figure}[htb!]
    \begin{minipage}[b]{1.0\linewidth}
      \centering
      \centerline{\includegraphics[width=16cm]{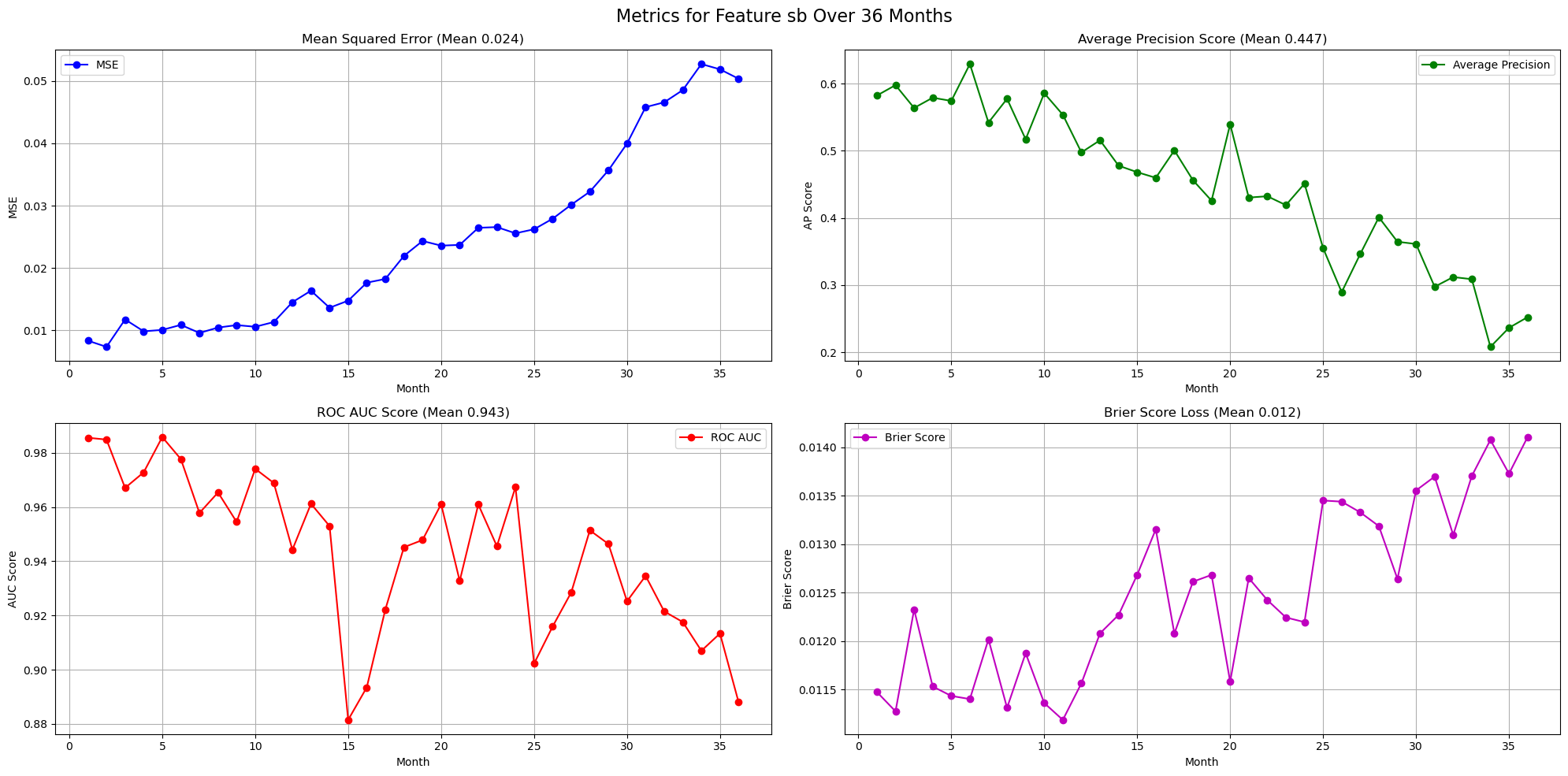}}
    \end{minipage}
    \caption{MSE, AP, AUC, and Brier scores for state-based violence across the full test partition of 36 months. This version includes the Middle East (ME) and is therefore not directly comparable with \cite{hegre2021views2020} and \cite{hegre2019views}. Predictions are based on the monthly mean of the sampled forecast distribution.
}\label{fig:sb_eval}
\end{figure}

\begin{figure}[htb!]
    \begin{minipage}[b]{1.0\linewidth}
      \centering
      \centerline{\includegraphics[width=16cm]{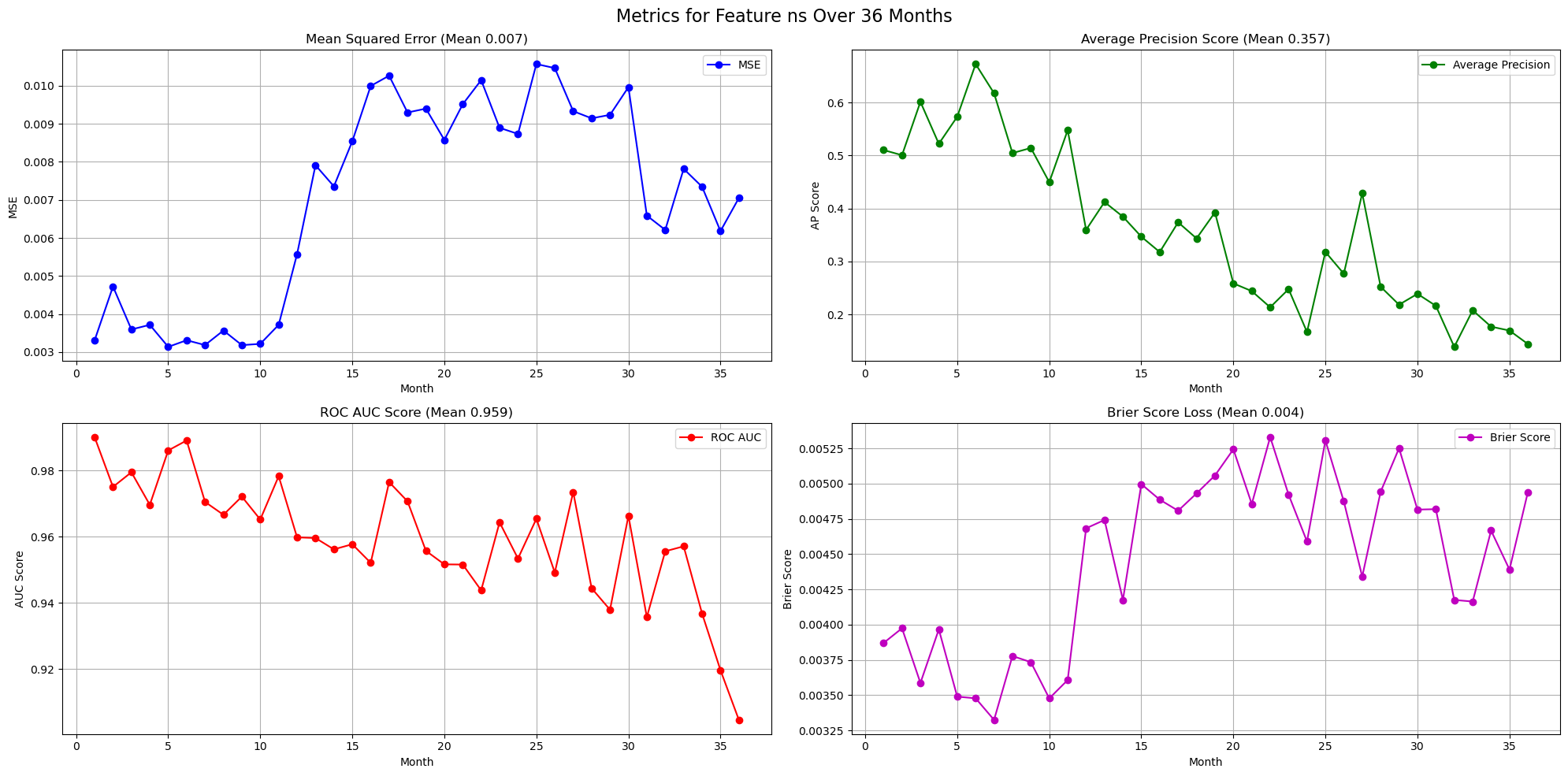}}
    \end{minipage}
    \caption{Evaluation metrics (MSE, AP, AUC, Brier) for non-state violence over 36 hold-out test months, using the full dataset including the Middle East (ME). As a result, the plot is not directly comparable with previous benchmarks \cite{hegre2021views2020, hegre2019views}. The monthly mean of the predictive distribution is used as the point forecast.
}\label{fig:ns_eval}
\end{figure}

\begin{figure}[htb!]
    \begin{minipage}[b]{1.0\linewidth}
      \centering
      \centerline{\includegraphics[width=18cm]{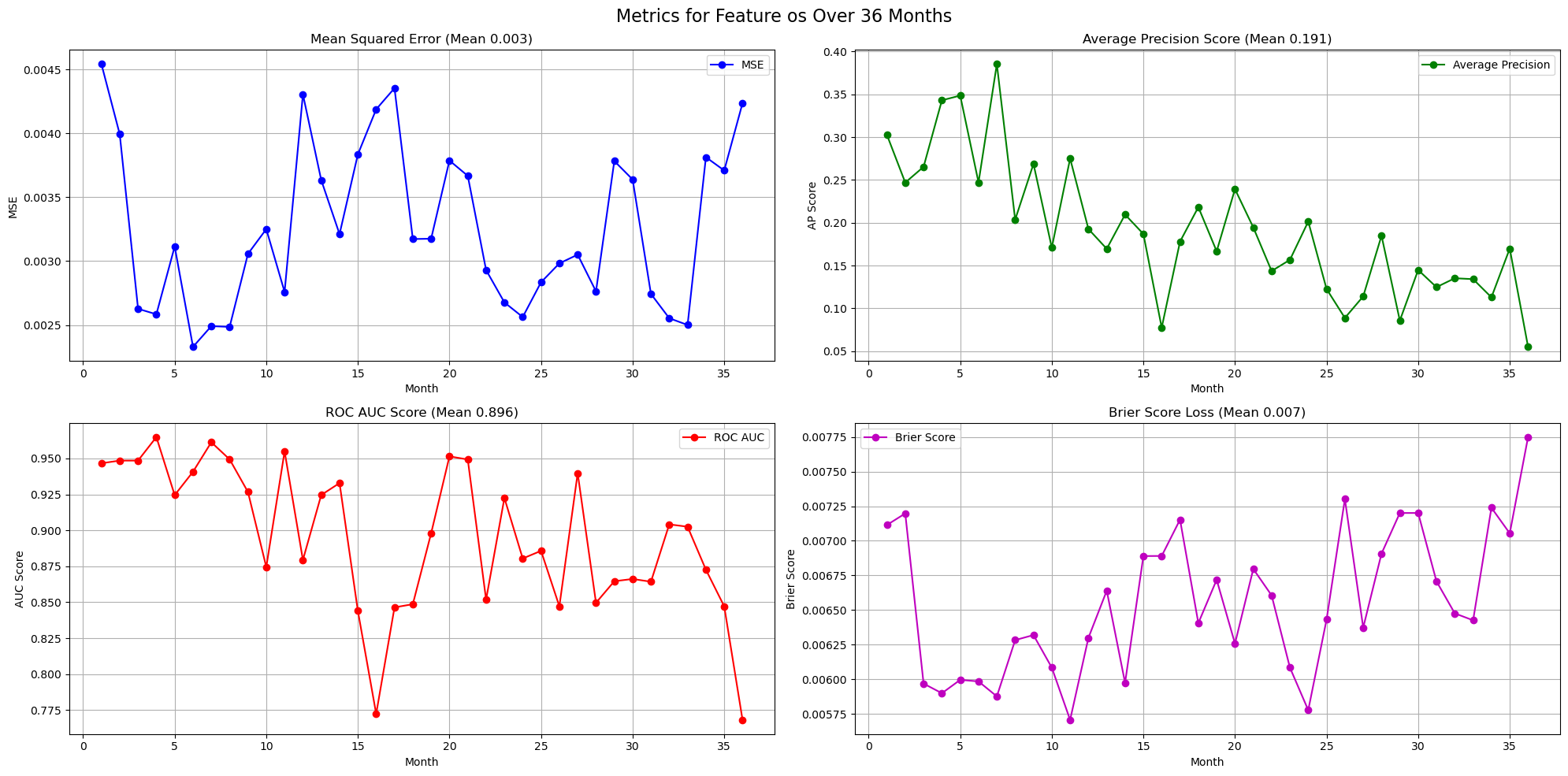}}
    \end{minipage}
    \caption{MSE, AP, AUC, and Brier scores for one-sided violence over 36 months of out-of-sample validation, with the Middle East (ME) region included. This makes the results not directly comparable to \cite{hegre2021views2020} and \cite{hegre2019views}. Predictive means are used as point estimates throughout.
}\label{fig:os_eval}
\end{figure}

%% file: tables/mean_results_me.tex
\begin{table}[ht!]
  \begin{center}
    \begin{tabular}{p{0.22\linewidth} p{0.22\linewidth}  p{0.22\linewidth} p{0.22\linewidth}}
    \toprule
        \textbf{Violence type}   & \textbf{Task/Metrics}               & \textbf{No-Change}             & \textbf{Hydra Net w/ ME}\\
        
        \midrule
        State-Based             
                                &  MSE                                  & 0.02                          & 0.024 \\
                                &  \textbf{AP}                           & \textbf{0.28}                  & \textbf{0.447} \\
                                &  AUC                                  & 0.70                          & 0.943 \\
                                &  Brier                                & 0.01                          & 0.012 \\
        \midrule
        Non-State-Based         
                                &  MSE                                 & 0.02                           & 0.007 \\
                                &  \textbf{AP}                          & \textbf{0.19}                   & \textbf{0.357} \\
                                &  AUC                                 & 0.65                           & 0.959 \\
                                &  Brier                               & 0.005                          & 0.004 \\
        \midrule
        One-sided               
                                &  MSE                                 & 0.02                           & 0.003 \\
                                & \textbf{AP}                           &  \textbf{0.08}     & \textbf{0.191} \\
                                &  AUC                                 & 0.63                           & 0.896 \\
                                &  Brier                               & 0.006                          & 0.007 \\
        \bottomrule
    \end{tabular}
    \caption{Evaluation scores for HydraNet (with the Middle East included) compared to a no-change benchmark, across three types of violence: state-based, non-state, and one-sided. Metrics include Mean Squared Error ($MSE$), Average Precision ($AP$), Area Under the Curve ($AUC$), and Brier score, calculated over the full 36-month test partition. As the Middle East is included, these results are not directly comparable with previous VIEWS benchmarks \citep{hegre2021views2020, hegre2019views}.}
\label{tab:mean_results_me}
  \end{center}
\end{table}